\documentclass[superscriptaddress,aps,preprintnumbers,showpacs,prd,nofootinbib,reprint]{revtex4-1}

\usepackage{graphicx} 
\usepackage{amsmath,amssymb,amsthm,slashed,bm}
\usepackage[colorlinks,allcolors=blue]{hyperref}

\usepackage{bm,epstopdf,natbib,hyperref,color,verbatim,multirow,bm,tikz,xcolor}
\hypersetup{colorlinks=true,urlcolor=blue,citecolor=blue,linkcolor=blue,menucolor=blue,anchorcolor=blue,filecolor=blue}
\everymath{\displaystyle}
\definecolor{lime}{HTML}{A6CE39}
\DeclareRobustCommand{\orcidicon}{
	\begin{tikzpicture}
	\draw[lime, fill=lime] (0,0) 
	circle [radius=0.16] 
	node[white] {{\fontfamily{qag}\selectfont \tiny ID}};
	\draw[white, fill=white] (-0.0625,0.095) 
	circle [radius=0.007];
	\end{tikzpicture}
	\hspace{-3mm}
}
\foreach \x in {A, ..., Z}{\expandafter\xdef\csname orcid\x\endcsname{\noexpand\href{https://orcid.org/\csname orcidauthor\x\endcsname}
			{\noexpand\orcidicon}}
}

\begin{document}

\preprint{TU-1323}

\title{Nuclear Recoils from Invisible Neutron-Pair Annihilation and the LZ Event}

\author{
Junseok Lee\hspace{-1mm}\orcidA{}
}
\email{lee.junseok.p4@dc.tohoku.ac.jp}
\affiliation{Department of Physics, Tohoku University, 
Sendai, Miyagi 980-8578, Japan}
\author{
Fuminobu Takahashi\hspace{-1mm}\orcidB{}
}
\email{fumi@tohoku.ac.jp}
\affiliation{Department of Physics, Tohoku University, 
Sendai, Miyagi 980-8578, Japan} 
\affiliation{Kavli IPMU (WPI), UTIAS, University of Tokyo, Kashiwa 277-8583, Japan}
\author{
Yu-Dai Tsai\hspace{-1mm}\orcidC{}
}
\email{y.tsai@sheffield.ac.uk, yudaitsai.academic@gmail.com}
\affiliation{School of Mathematical and Physical Sciences, University of Sheffield, Sheffield S3 7RH, UK}
\affiliation{The University of Manchester, Manchester M13 9PL, UK}

\date{September 10, 2026}

\begin{abstract}
Two bound neutrons can annihilate into a single invisible scalar carrying baryon number two, producing a monochromatic nuclear recoil. Pair removal connects the $0^+$ ground states of even-even nuclei and, for several detector isotopes, leaves a stable daughter where single-neutron removal would leave a radioactive one.
Below the first daughter excitation, the leading transition produces neither nuclear de-excitation nor subsequent daughter decay.
We identify the corresponding mass windows in argon and xenon, including argon recoils up to $108\,\mathrm{keV}$, and obtain an approximate partial-lifetime bound of $8.5\times10^{26}\,\mathrm{yr}$ from DEAP-3600 data. 
We also discuss this mechanism as a possible origin of the
$248\,\mathrm{keV}$ recoil candidate reported by LZ.
A ground-state interpretation points to a scalar mass near
$1.85\,\mathrm{GeV}$, where daughter excitations are also accessible, and motivates searches for correlated higher-energy recoil lines in argon.
Recoil lines in different isotopes reconstruct a common invisible-particle mass.
The emitted scalar can itself be dark matter, making these searches a probe of ordinary matter converting into a dark sector.
\end{abstract}

\maketitle

\textit{\textbf{Introduction.}}---
The search for dark matter has driven the development of detectors capable of observing rare nuclear recoils in large target masses~\cite{DEAP:2019yzn,LZ:2026axp,XENON:2025vwd,PandaX:2024qfu}. While such recoils are usually associated with the scattering of an incident particle, they can also originate from a transition within the nucleus. Emission of an invisible particle would cause the daughter nucleus to recoil, providing a direct probe of new interactions between ordinary matter and a dark sector.

Invisible nucleon decay has been investigated in particle models~\cite{Helo:2025kgx} and in searches for nuclear de-excitation products or radioactive daughters~\cite{KamLAND:2005pen,SNO:2022trz,Borexino:2003igu}. These signatures can be absent when the transition leaves a stable daughter in its ground state. In that case, the nuclear recoil itself offers a complementary search channel.

In this Letter, we consider the conversion of two bound neutrons into a single scalar particle $\phi$ carrying baryon number two,
\begin{equation}
 (A,Z)\longrightarrow(A-2,Z)^{(*)}+\phi .
 \label{eq:process}
\end{equation}
The daughter recoils against $\phi$ at an energy fixed by two-body kinematics. In argon and xenon, there are mass intervals in which only a stable daughter ground state can be populated, giving a single nuclear-recoil line. We identify these intervals and the correlated line energies in different isotopes. Using published liquid-argon data from DEAP-3600, we illustrate the sensitivity to the nuclear conversion lifetime. 
Below these intervals, daughter excitations become accessible and higher recoil energies are possible. We discuss the corresponding nuclear response, with particular relevance to  the $248 \pm 23\,\mathrm{(stat)} \pm 23\,\mathrm{(sys)}\,\,\mathrm{keV}$ recoil candidate reported by LUX-ZEPLIN (LZ) in $2.84$ tonne years~\cite{LZ:2026axp}.
The LZ event has already motivated a number of theoretical studies based on dark matter~\cite{Su:2026rwz,Fan:2026kxx,Freese:2026sga,Wu:2026nhi,Lou:2026idn,Yin:2026jnn,Nomura:2026qyq,DiMauro:2026ldr,Pospelov:2026ewn,Visinelli:2026kgt,Yamashita:2026ump,Smirnov:2026aqk,Du:2026guj,Rodd:2026tyn,McCabe:2026crm,Unwin:2026rdp,Dent:2026bji,deLima:2026shq,Gu:2026vto,Baer:2026fpy,Lee:2026wof,Wang:2026ytg,Yang:2026wpb,Kotlarski:2026pep,Liang:2026coz,DiMauro:2026dqp,Das:2026uyy,Alhazmi:2026efz,Okada:2026eol,Ahmed:2026qjg,Du:2026lpa,Bandyopadhyay:2026gjw,Borah:2026zwf,Bose:2026ndd,Bisal:2026khf,Khan:2026nwp,Cheung:2026byg,Yuan:2026djt,Elahi:2026vlm,Zhu:2026dag,Asadi:2026iot,Lee:2026xxh,Lee:2026jxl,Langhoff:2026ujr}, boosted dark particles~\cite{Kannike:2026qyl} and neutrinos~\cite{Jeesun:2026vzo}.

\textit{\textbf{A neutron-pair interaction.}}---Take a neutral complex scalar with baryon number $B_\phi=2$ and positive parity. A leading nucleon interaction is
\begin{equation}
    \mathcal{L}_{\rm int}
    =
    y \phi^\dagger n^T C \gamma_5 n + \mathrm{h.c.} \, ,
    \qquad
    \langle \phi \rangle = 0 .
    \label{eq:interaction}
\end{equation}
Here $n$ is a Dirac neutron field, $y$ is a dimensionless coupling and $C=i\gamma^2\gamma^0$ is the
charge-conjugation matrix.
In the Dirac representation, the leading nonrelativistic
interaction is
\begin{equation}
 \begin{aligned}
 \mathcal L_{\rm int}^{\rm NR}
 &=-y\phi^\dagger\psi^T i\sigma_2\psi+\mathrm{h.c.}\\
 &=2y\phi^\dagger\psi_\downarrow\psi_\uparrow
   +\mathrm{h.c.},
 \end{aligned}
 \label{eq:interaction_NR}
\end{equation}
where $\psi=(\psi_\uparrow,\psi_\downarrow)^T$ is the two-component neutron annihilation field, with its rest-energy phase retained, and $\phi$ retains its relativistic normalization.
This operator removes a spin-singlet neutron pair in a relative $S$ wave, the ${}^{1}S_0$ channel.

The interaction conserves total baryon number, transferring two units from ordinary matter to $\phi$.
With baryon number conserved in the low-energy theory and $\langle\phi\rangle=0$, it generates no vacuum neutron--antineutron mass mixing.
The mass range of interest, $m_\phi \simeq 1.85$--$1.87\,\mathrm{GeV}$, also forbids emission of $\phi$ by an isolated nucleon. At the quark level, such a coupling can arise, for example, from gauge-invariant dimension-ten operators, schematically $\phi^\dagger(u_Rd_Rd_R)^2/\Lambda^6$, with $\Lambda$ a heavy mass scale and Lorentz and color contractions understood. Electric charge selects neutron pairs for the neutral two-to-one transition. We assume that the emitted scalar escapes the nucleus and the detector without depositing observable energy.

\textit{\textbf{Ground-state transitions.}}---The $^{1}S_0$ pair interaction can connect the $0^+$ ground states of even-even nuclei. Argon and xenon provide useful examples: pair conversion in $^{40}\mathrm{Ar}$ and $^{132,134,136}\mathrm{Xe}$ leaves stable $^{38}\mathrm{Ar}$ and $^{130,132,134}\mathrm{Xe}$, respectively~\cite{ENSDF}. When the available energy lies below the first daughter excitation and all breakup thresholds, only the ground-state transition is allowed, yielding a single nuclear-recoil line.

Let $M_A$ and $M_D$ be the parent and daughter ground-state masses, and define $\Delta_A = M_A - M_D = 2m_n - S_{2n,A}$ and $Q_A = \Delta_A - m_\phi$, where $S_{2n,A}$ is the two-neutron separation energy.
For a daughter nuclear state $f$ with excitation energy $E_f^*$,
exact two-body kinematics gives the daughter recoil energy,, exact two-body kinematics gives the daughter recoil energy,
\begin{equation}
    T_{A,f} = \frac{(\Delta_A - E_f^*)^2 - m_\phi^2}{2M_A}\,,
    \qquad
    0\leq E_f^*<Q_A.
    \label{eq:recoil}
\end{equation}
Near threshold, $T_{A,f} \simeq (m_\phi / M_A)(Q_A - E_f^*)$: a MeV of available energy gives about $15\,\mathrm{keV}$ in xenon or $50\,\mathrm{keV}$ in argon.
We use neutral ground-state atomic masses~\cite{AME}
to evaluate the reference line energies.
These energies correspond to transitions between atoms in their electronic ground states;
possible electronic excitation or ionization would modify the detector response
and is not included here.
For a fixed daughter level, the recoil energy is fixed by the parent and daughter masses and $m_\phi$,
while nuclear structure determines the transition strength.

Let $E_{{\rm gap},A}$ be the minimum
of the first daughter excitation energy and the light-particle breakup thresholds. The nuclear ground state is the only accessible channel when
\begin{equation}
 \Delta_A-E_{{\rm gap},A}<m_\phi<\Delta_A .
 \label{eq:gap}
\end{equation}
For a stable daughter, the leading transition in this interval produces a nuclear recoil unaccompanied by nuclear de-excitation or later radioactive decay. Table~\ref{tab:windows} lists the corresponding mass windows and maximum recoil energies; the first daughter excitation sets each lower boundary.

\begin{table}[!t]
    \caption{
        Ground-state-only pair-conversion windows from evaluated masses and levels~\cite{AME,ENSDF}. All listed daughters have $0^+$ ground states and are stable on detector timescales. The last column gives the maximum recoil energy within each open mass interval.
    }
    \label{tab:windows}
    \begin{ruledtabular}
        \begin{tabular}{ccc}
            Parent $\rightarrow$ daughter & $m_\phi$ interval [MeV] & $T_{\max}$ [keV] \\
            $^{40}$Ar $\rightarrow {}^{38}$Ar & $(1860.496, 1862.663)$ & $108.39$ \\
            $^{132}$Xe $\rightarrow {}^{130}$Xe & $(1863.054, 1863.590)$ & $8.13$ \\
            $^{134}$Xe $\rightarrow {}^{132}$Xe & $(1863.474, 1864.141)$ & $9.98$ \\
            $^{136}$Xe $\rightarrow {}^{134}$Xe & $(1863.838, 1864.685)$ & $12.47$ \\
        \end{tabular}
    \end{ruledtabular}
\end{table}

\textit{\textbf{Recoil signatures.}}---Argon is especially useful because the first $^{38}$Ar excitation is at $2.1675\,\mathrm{MeV}$~\cite{ENSDF}. Thus an isolated nuclear recoil can extend to $108.4\,\mathrm{keV}$ without any accessible daughter excitation. At $m_\phi=1861\,\mathrm{MeV}$, $^{40}$Ar produces an $83.2\,\mathrm{keV}$ recoil and a stable ground-state daughter.

\begin{figure}[t]
    \includegraphics[width=\columnwidth]{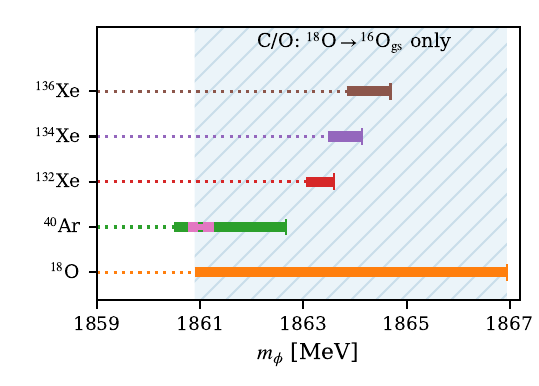}
    \caption{
        Mass intervals for neutron-pair conversion in argon and xenon, obtained from evaluated masses and nuclear levels~\cite{ENSDF,AME}.
        For each isotope, dotted segments denote the region in which conversion is kinematically allowed and excited daughter states are accessible, while thick segments mark the ground-state-only interval, $\Delta_A-E_{{\rm gap},A}<m_\phi<\Delta_A$.
        The vertical tick at the right end marks the conversion endpoint $m_\phi=\Delta_A$; the channel is closed for heavier $\phi$.
        The shaded band, $1860.893<m_\phi<1866.942~\mathrm{MeV}$, refers to natural carbon and oxygen:
        the $^{12,13}\mathrm{C}$ and $^{16,17}\mathrm{O}$ channels are closed, leaving only $^{18}\mathrm{O}\to{}^{16}\mathrm{O}_{\rm gs}+\phi$, which produces neither nuclear de-excitation nor a radioactive daughter.
        Squares on the $^{40}\mathrm{Ar}$ row delimit the illustrative DEAP interval corresponding to $75<T_{{}^{40}{\rm Ar}}<90~\mathrm{keV}$.
    }
    \label{fig:windows}
\end{figure}

Carbon and oxygen, used in invisible nucleon-decay searches, illustrate the complementarity of the recoil signal. For
\begin{equation}
    1860.893<m_\phi<1866.942\,\mathrm{MeV},
    \label{eq:COwindow}
\end{equation}
conversion is closed in $^{12,13}$C and $^{16,17}$O~\cite{AME}. Only $^{18}\mathrm{O} \rightarrow {}^{16}\mathrm{O}_\mathrm{gs}+\phi$ remains open, where the subscript $\mathrm{gs}$ denotes the ground state. 
The first $^{16}\mathrm{O}$ excitation, at $6.0494\,\mathrm{MeV}$, is inaccessible~\cite{ENSDF}. The parent isotope ${}^{18}\mathrm{O}$, with a natural abundance of approximately $0.2\%$~\cite{CIAAW}, can produce an oxygen recoil but neither nuclear de-excitation nor daughter radioactivity. Figure~\ref{fig:windows} shows how the argon and xenon recoil windows overlap this interval.

Natural xenon has its own fully ground-state-only interval, $1863.838 < m_\phi < 1864.685\,\mathrm{MeV}$.
All xenon channels with $A\leq132$, including odd nuclei,
are then closed; only $^{134,136}$Xe can contribute, and above $1864.141\,\mathrm{MeV}$ only $^{136}$Xe remains.
For example, $m_\phi=1864.4\,\mathrm{MeV}$ gives a $4.20\,\mathrm{keV}$ recoil.
More generally, two correctly assigned ground-state lines from isotopes $A$ and $B$ must reconstruct the same mass:
\begin{equation}
    \Delta_A^2 - 2 M_A T_A = \Delta_B^2 - 2 M_B T_B = m_\phi^2,
    \label{eq:mass}
\end{equation}
where $T_A \equiv T_{A,\rm{gs}}$ and similarly for $B$.
For $^{134,136}$Xe this is $T_{{}^{134} {\rm Xe}} \simeq 1.01495 T_{{}^{136} {\rm Xe}} - 8.124\,\mathrm{keV}$.
In the fully ground-state-only interval, $T_{{}^{134} {\rm Xe}}<4.54\,\mathrm{keV}$, so observing both lines requires sufficiently low thresholds. Argon and xenon cover separate ground-state-only mass intervals, as shown in Fig.~\ref{fig:windows}.

Varying the isotopic composition provides a further test.
Enriching a particular isotope strengthens its recoil line
without changing its energy.
After accounting for exposure and detection efficiency,
the line strength increases in proportion to the isotope fraction.

\textit{\textbf{An existing recoil constraint.}}---The absence of a recoil line bounds the partial lifetime for conversion to the daughter ground state. To relate this observable to the microscopic interaction, define the dimensionless pair-removal amplitude
\begin{equation}
  F_{A,f}(\mathbf p)=\left\langle f\left|\int d^3x\,e^{-i\mathbf p\cdot\mathbf x}\psi_\downarrow(\mathbf x)\psi_\uparrow(\mathbf x)\right|A\right\rangle .   
\end{equation}
Here $|A\rangle$ is the parent ground state and $|f\rangle$
is the daughter state.
The overall center-of-mass motion is factored out, and internal nuclear states have unit norm. With the coupling $2y$ in Eq.~\eqref{eq:interaction_NR}, the leading nonrelativistic nuclear source gives the partial conversion rate to state $f$
\begin{equation}
    \Gamma_{A,f}=\frac{2|y|^2}{\pi}p_f\frac{E_{D,f}}{M_A}
    \overline{|F_{A,f}(p_f)|^2} .
    \label{eq:rate}
\end{equation}
Here $p_f$ is the emitted momentum, $E_{D,f}$ the total daughter energy, and the bar averages over the initial spin and emission direction and sums over the final spin. The nonrelativistic pair coupling is $2y$ in this convention. Each isotope has its own pair response. For a nonzero $F_{A,\mathrm{gs}}(0)$, a $0^+ \rightarrow 0^+$ rate vanishes as $\sqrt{Q_A}$ at threshold.

The DEAP-3600 search observed no events in its signal region in $230.63$ live days, with fiducial mass $824\,\mathrm{kg}$~\cite{DEAP:2019yzn}.
For illustrative purposes, we consider representative line energies of $75$--$90\,\mathrm{keV}$.
We denote by $\epsilon_A$ the total acceptance for the
ground-state recoil line from parent isotope $A$.
We approximate the response to the recoiling $^{38}\mathrm{Ar}$
daughter by that to ordinary argon nuclear recoils, taking
$\epsilon_{{}^{40}\mathrm{Ar}}\simeq 0.25$ over this recoil-energy range~\cite{DEAP:2020iwi}.
The expected count is $N_{{}^{40}{\rm Ar}}\,\Gamma_{{}^{40}{\rm Ar},\mathrm{gs}}\, t\, \epsilon_{{}^{40}{\rm Ar}}$, where $N_{{}^{40}\mathrm{Ar}}=1.24\times10^{28}$ is the number
of ${}^{40}\mathrm{Ar}$ nuclei in the fiducial volume
and $t$ is the live time.
Treating the exposure and acceptance as fixed and subtracting no background gives the one-sided Poisson upper count $s_{90}=\ln10$, hence the approximate $90\%$ confidence bound
\begin{equation}
\Gamma_{{}^{40}{\rm Ar},\mathrm{gs}}^{-1}
    \gtrsim
    8.5 \times 10^{26} \,\mathrm{yr} \left( \frac{\epsilon_{{}^{40}{\rm Ar}}}{0.25} \right)
    .
    \label{eq:DEAP}
\end{equation}
This is a partial lifetime per parent nucleus and applies to $1860.864$--$1861.164\,\mathrm{MeV}$, inside the argon ground-state-only window.
At $80\,\mathrm{keV}$, Eq.~\eqref{eq:rate} gives
\begin{equation}
    |y F_{{}^{40}{\rm Ar}}|
    \lesssim
    2.3 \times 10^{-29} \left( \frac{0.25}{\epsilon_{{}^{40}{\rm Ar}}} \right)^{1/2}.
\end{equation}
Here $F_{{}^{40}{\rm Ar}}$ is the ground-state amplitude at the corresponding momentum, $p_f\simeq 75\,\mathrm{MeV}$. Extracting a bound on $y$ separately requires a nuclear calculation.

\textit{\textbf{The higher-energy LZ candidate.}}---Assigning the $248\,\mathrm{keV}$ recoil to $^{132}\mathrm{Xe} \rightarrow {}^{130}\mathrm{Xe}_{\rm gs}+\phi$ instead gives $m_\phi=1847.17\,\mathrm{MeV}$, $Q_{{}^{132}{\rm Xe}}=16.42\,\mathrm{MeV}$, and $p_f=245\,\mathrm{MeV}$.
Combining the quoted recoil-energy uncertainties in quadrature gives an uncertainty $\sigma_m\simeq 2.2\,\mathrm{MeV}$ for this isotope assignment.
For the $2.84$ tonne-year exposure~\cite{LZ:2026axp}, 
one expected accepted ground-state event corresponds to $\sum\nolimits_A\eta_A\epsilon_A\Gamma_{A,\mathrm{gs}}=7.7\times10^{-29}\,\mathrm{yr}^{-1}$, 
where $\eta_A$ is the isotope number fraction.

The ground-state line is the endpoint of the primary recoil spectrum for this isotope. Transitions to excited $^{130}$Xe states give smaller primary recoils and can be accompanied by nuclear de-excitation products.
Since $Q_{{}^{132}{\rm Xe}}$ exceeds the neutron separation energy $S_n(^{130}\mathrm{Xe})=9.26\,\mathrm{MeV}$~\cite{AME}, both electromagnetic de-excitation and neutron-emitting branches are accessible. Their observed rates depend on the pair-removal amplitudes, daughter decay branching fractions, and detector response. Events accompanied by prompt de-excitation photons or
neutron-induced activity may be rejected by the veto or other
event-selection cuts.
The ground-state recoil line can therefore dominate the
accepted signal even if its branching fraction is small.
The larger total conversion rate needed to compensate for such
a small branching fraction can still be compatible with the
light-nucleus bounds discussed below.

The strong ground-state transition observed in the two-neutron transfer reaction $^{138}\mathrm{Ba}(p,t){}^{136}\mathrm{Ba}$~\cite{Rebeiro:2020wvo} illustrates that neutron-pair removal can leave an even-even daughter unexcited. The transfer reaction and the local operator in Eq.~\eqref{eq:interaction_NR} weight the nuclear wavefunctions differently, so the xenon branching fractions require a separate evaluation at the relevant emitted momenta.

Independent constraints arise from the same conversion in lighter nuclei. At the central value $m_\phi=1847.17\,\mathrm{MeV}$, the available energies for 
$^{12}\mathrm{C}\to{}^{10}\mathrm{C}+\phi$ and $^{16}\mathrm{O}\to{}^{14}\mathrm{O}+\phi$ are $0.12$ and $3.08\,\mathrm{MeV}$, respectively. The carbon channel can therefore open or close within the mass uncertainty inferred from the LZ event.
When these channels are open, the radioactive daughters searched for by the BOREXINO Counting Test Facility (CTF) imply the approximate $90\%$ bounds $\Gamma_{{}^{12}{\rm C},\mathrm{gs}}^{-1}>2.2\times10^{25}\,\mathrm{yr}$ and $\Gamma_{{}^{16}{\rm O},\mathrm{gs}}^{-1}>5.6\times10^{24}\,\mathrm{yr}$~\cite{Borexino:2003igu}.
For a benchmark in which $^{132}$Xe alone supplies one expected accepted event, its $26.9\%$ natural abundance~\cite{CIAAW} implies $\Gamma_{{}^{132}\mathrm{Xe},\mathrm{gs}}^{-1}\simeq3.5\times10^{27}\epsilon_{{}^{132}{\rm Xe}}\,\mathrm{yr}$. The CTF bounds then require
\begin{equation}
    \begin{aligned}
        \frac{\Gamma_{12\mathrm{C},\mathrm{gs}}}{\Gamma_{{}^{132}{\rm Xe},\mathrm{gs}}}
        &\lesssim1.6\times10^2\epsilon_{{}^{132}{\rm Xe}},\\
        \frac{\Gamma_{16\mathrm{O},\mathrm{gs}}}{\Gamma_{{}^{132}{\rm Xe},\mathrm{gs}}}
        &\lesssim6.3\times10^2\epsilon_{{}^{132}{\rm Xe}}.
    \end{aligned}
    \label{eq:CTF_ratios}
\end{equation}
The common coupling cancels in these ratios through Eq.~\eqref{eq:rate}, leaving conditions on the relative nuclear responses.
Even if the ground-state branching fraction is small, these bounds
can leave room for a compensating increase in the total xenon
conversion rate.

Radioactive daughters in odd-mass xenon provide a related internal cross-check; for example, $^{129}$Xe pair removal produces $^{127}$Xe, whose subsequent decay has been searched for by DAMA/LXe~\cite{Bernabei:2000xp}.
Finally, at the same $m_\phi$ the $^{40}$Ar ground-state recoil is about $0.77\,\mathrm{MeV}$, well above the DEAP energy interval used above. The DEAP limit therefore probes a different scalar-mass region, while a dedicated higher-energy argon search could test the LZ-motivated mass directly.

\textit{\textbf{Discussion.}}---Neutron-pair annihilation connects even-even nuclei to effectively stable daughters through their $0^+$ ground states. Near threshold, a nuclear-recoil line can be the only signal of the leading transition.
The argon and xenon mass windows, the DEAP partial-lifetime bound, and the common mass reconstructed from isotope lines provide concrete targets for recoil searches.
The conversion rate is intrinsic to the target nucleus and independent of the ambient dark-matter density.
At higher recoil energies, accessible excited-state branches
provide additional tests through nuclear de-excitation and
daughter decays.
For example, de-excitation products from the excited
${}^{134}\mathrm{Xe}$ daughter of ${}^{136}\mathrm{Xe}$ conversion
could be sought in KamLAND-Zen~\cite{KamLAND-Zen:2024eml},
which uses a ${}^{136}\mathrm{Xe}$-enriched target.

The emitted scalar is a dark dibaryon and can also constitute dark matter.
Its mass lies below the neutral deuterium threshold near $1876\,\mathrm{MeV}$~\cite{AME}, so baryon-number conservation forbids decay into ordinary matter. It is therefore stable in the absence of lighter dark decay channels.
Its baryon charge and GeV-scale mass motivate an asymmetric origin of its abundance~\cite{Kaplan:2009ag,Davoudiasl:2010am}. For an asymmetry-dominated relic, $\Omega_\phi/\Omega_b\simeq[m_\phi/(2m_n)]|Y_B^{\rm dark}|/Y_B^{\rm vis}$, where $Y_B^{\rm dark}=2(n_\phi-n_{\bar\phi})/s$ and $Y_B^{\rm vis}$ are the dark and visible baryon-charge asymmetries per entropy density $s$. Since $m_\phi/2\simeq m_n$, comparable charge asymmetries give comparable energy densities. The six-quark interaction offers a possible route for sharing a primordial asymmetry between the two sectors. A detailed study of the resulting relic abundance and dark-matter--baryon density ratio, including asymmetry transfer and depletion of the symmetric component, is left for future work. In such a realization, the recoil would signal the conversion of ordinary baryon number into dark matter.

\textit{\textbf{Other physical implications.}}---
Conversion of neutron pairs inside neutron stars could provide further constraints, independent of the ambient dark matter assumptions.
A thermalized bosonic population could condense and soften the equation of state~\cite{McKeen:2018xwc,Shahrbaf:2024gdm}.
Potential constraints and effects require dedicated calculations, together with a specification of the $\phi$ self-interactions and its interactions with nuclear matter.

As an interesting extension to the model, we can consider a common source that produces equal asymmetries in $\phi$ particle number and visible sector $B-L$, $|Y_\phi| = |Y_{B-L}^{\mathrm{visible}}|$, before electroweak sphaleron freeze-out.
Here $Y_\phi\equiv(n_\phi-n_{\bar\phi})/s=Y_B^{\rm dark}/2$,
and $Y_{B-L}^{\rm visible}$ is the visible-sector $B-L$
asymmetry per entropy density.
For example, CP-asymmetric decays of heavy neutral fermions into $\phi + udd$ and the conjugate final state would generate this relation through their final state particle content.
If subsequent transfer and washout are negligible, and if $\phi$ constitutes all of the dark matter, the Standard Model sphaleron conversion, $Y_B^{\mathrm{visible}} = (28/79)\times\, Y_{B-L}^{\mathrm{visible}}$~\cite{Harvey:1990qw}, together with the observed dark-
matter–baryon density ratio,  predict the mass of $\phi$ to be about $1.8$\,GeV.
This happens to be close to the mass range suggested by our scenario.
This numerical coincidence warrants further investigation.

In addition, so far we have considered an elementary dark dibaryon, but it could be interesting to consider a composite dibaryon made of the Standard Model quarks, the hypothetical $J^P = 0^+$ sexaquark $S \sim uuddss$~\cite{Moore:2024mot}. When the mass of the sexaquark is around $1878\,\mathrm{MeV}$, the main decay channels are closed~\cite{Moore:2024mot}. 
The recoil kinematics also applies to a single composite $B=2$ state.
In this case, $n n \rightarrow S$ changes strangeness by two units and requires weak interactions in the Standard Model.
Compositeness preserves Eq.~(\ref{eq:interaction_NR}) for a specified two-body final state.
However, for the event to register as an isolated nuclear recoil, the sexaquark must escape the daughter nucleus without significant rescattering, which requires further study.

\vspace{2mm}

\begin{acknowledgments}
We thank Yasuhiro Kishimoto for helpful discussions on possible
experimental signatures at KamLAND-Zen.
This work was supported by JSPS KAKENHI Grant Numbers 25H02165 (F.T.), 26K00695 (F.T.), and 25KJ0564 (J.L.), by the World Premier International Research Center Initiative (WPI), MEXT, Japan, and by COST Action COSMIC WISPers CA21106. J.L. was also supported by the Graduate Program on Physics for the Universe (GPPU), Tohoku University. Y.-D.T. is supported by a Dorothy Hodgkin Fellowship funded by the Royal Society and acknowledges start-up support from the University of Sheffield.
The results in this draft were cross-checked, and the grammar was polished using advanced tools.
\end{acknowledgments}

\vspace{3mm}

\textit{\textbf{Note added.}}---While this work was being completed, Ref.~\cite{AghaieStrumia} appeared, studying single-neutron disappearance in connection with the LZ event through spontaneous decay and dark-matter-induced reactions.

\bibliographystyle{apsrev4-1}
\bibliography{references}

\end{document}